\documentclass[12pt]{article}
\usepackage{graphicx}
\usepackage{latexsym, pstricks}
\usepackage{amsmath,amssymb,amsthm}
\usepackage[labelformat=empty,labelsep=none]{caption}

\font\titulo=cmbx10 scaled\magstep1 
scaled\magstep1

\def\section#1{\vskip 1.5truepc plus 0.1truepc minus 0.1truepc
    \goodbreak \leftline{\titulo#1} \nobreak \vskip 0.1truepc
    \indent}
\def\frc#1#2{\leavevmode\kern.1em
    \raise.5ex\hbox{\the\scriptfont0 $ #1 $}\kern-.1em
    /\kern-.15em\lower.25ex\hbox{\the\scriptfont0 $ #2 $}}

\def\IZ{{\rm Z}\llap{\vrule height7.1pt width1pt
     depth-.4pt\phantom t}} 

\newbox\pmbbox
 \def\pmb#1{{\setbox\pmbbox=\hbox{$#1$}%
\copy\pmbbox\kern-\wd\pmbbox\kern.3pt\raise.3pt\copy\pmbbox\kern-\wd\pmbbox
\kern.3pt\box\pmbbox}}

\font\cmss=cmss10 \font\cmsss=cmss10 at 7pt

\def\IZ{\relax\ifmmode\mathchoice
{\hbox{\cmss Z\kern-.4em Z}}{\hbox{\cmss Z\kern-.4em Z}}
{\lower.9pt\hbox{\cmsss Z\kern-.4em Z}} {\lower1.2pt\hbox{\cmsss
Z\kern-.4em Z}}\else{\cmss Z\kern-.4em Z}\fi}

\font\cmss=cmss10 \font\cmsss=cmss10 at 7pt
\def\IS{\relax\ifmmode\mathchoice
{\hbox{\cmss S\kern-.4em S}}{\hbox{\cmss S\kern-.4em S}}
{\lower.9pt\hbox{\cmsss S\kern-.4em S}} {\lower1.2pt\hbox{\cmsss
S\kern-.4em S}}\else{\cmss S\kern-.4em S}\fi}

\begin{document}

\centerline{\titulo The Role of Boolean Irreducibility}

\centerline{\titulo  in \textit{NK}-Kauffman Networks}

\vskip 1.0pc \centerline{Federico Zertuche}

\vskip 1.0pc \centerline{Instituto de Matem\'aticas, Unidad
Cuernavaca} \centerline{Universidad Nacional Aut\'onoma de
M\'exico} \centerline{A.P. 273-3, 62251 Cuernavaca, Morelos,
M\'exico.} \centerline{\tt federico.zertuche@im.unam.mx}

\vskip 1.5pc {\bf \centerline {Abstract}}

Boolean variables are such that they take only values on $ \mathbb{Z}_2 \cong \left\{ 0, 1 \right\} $. \textit{NK}-Kauffman networks are dynamical deterministic systems of $ N $ Boolean functions that depend only on $ K \leq N $ Boolean variables. They were proposed by Kauffman as a first step to understand cellular behaviour [Kauffman, S.A.; {\rm The Large Scale Structure and Dynamics of Gene Control Circuits: An Ensemble Approach}. {\it J.~Theoret.~Biol.} {\bf 44} (1974) 167.] with great success. Among the problems that still have been not well understood in Kauffman networks, is the mechanism that regulates the phase transition of the system from an ordered phase; where small changes of the initial state decay, to a chaotic, where they grow exponentially. We show, that this mechanism is regulated through \textsf{the irreducible decomposition} of Boolean functions proposed in [ Zertuche, F. {\rm On the robustness of NK-Kauffman networks against changes in their connections and Boolean functions}, {\it J.~Math.~Phys.} {\bf 50} (2009) 043513]. This is in contrast to previous knowledge that attributed it to \textsf{canalization}. We also review other statistical properties of Kauffman networks that have been shown that \textsf{Boolean irreducibility} explains.

\vskip 1.0pc

\noindent {\bf Short title:} {\it The Role of Boolean Irreducibility}

\vskip 0.5pc \noindent {\bf Keywords:} {\it Cellular automata,
Boolean irreducibility, binary functions, phase transitions,
\textit{NK}-Kauffman Networks}.


\newpage

\baselineskip = 12.4pt

\section{1. Introduction.}

\textit{NK}-Kauffman networks were proposed by Stuart A. Kauffman in 1969 as a starting point of a mechanism that mimics the transition from disorder to order in living organisms~${}^{1,2}$. Nevertheless its simplicity in construction a wide literature, in theoretical biology, statistical mechanics, and also pure mathematics; has been dedicated to the subject with many extensions to the original model. See Refs.~[2,3] and references therein.

A \textit{NK}-Kauffman network consists of a set of $ N $ Boolean variables \break $ S_i (t) \in \mathbb{Z}_2 $ ($ i = 1, \dots, N $), which evolve deterministically and synchronously (all the variables simultaneously) in a discrete time $ t = 0, 1, 2, \dots $ This is done according to $ N $, in general different, $K$-Boolean functions
$$
b_K^{(i)}: \mathbb{Z}_2^K \to \mathbb{Z}_2
$$
that depend only on $ K $ ($ 0 \leq K \leq N $) of the $ N $ variables at the previous time. Concretely, the evolution rule is given at each time step $t$ by
$$
S_i (t+1) = b_K^{(i)} \left( S_{i_1}(t), S_{i_2}(t), \dots, S_{i_K}(t)
\right), \ \ i = 1, \dots, N, \eqno(1)
$$
where each the $K$-Boolean functions $ b_K^{(i)} $ is chosen randomly and independently with a bias probability
$$
0 < p < 1 , \eqno(2)
$$
that $ b_K^{(i)} = 1 $; for each of its possible $ 2^K $ arguments, and $ b_K^{(i)} = 0 $ with probability $ 1 - p $. For each site $ i $, $ K $ inputs (called the connections) are randomly selected with equiprobability among the $ N $ Boolean variables of the network, without repetition. Once this random selection has been done a Boolean dynamically deterministic \textit{NK}-Kauffman network has been, randomly, constructed. Note however, that its subsequent dynamics is deterministic, according to (1), and it will depend only on which state $ {\bf S} \left( 0 \right) \in \mathbb{Z}_2^N $ the evolution starts. The goal now is to consider the set $ {\cal L}^N_K $ of \textit{NK}-Kauffman networks and try to understand the average dynamics of their elements under the extraction rule (2). Since the number of states $ {\bf S} = \left( S_1, \ldots, S_N \right) \in \mathbb{Z}_2^N $ is finite ($ 2^N $), the dynamics eventually settles down onto a cycle. Note also that, depending on which state the network starts, its dynamics may end on other cycle. All the states $ {\bf S} $ may be put in a one-to-one correspondence between the first $ 2^N $ naturals. This is done through the bijection
$$
s \left( {\bf S} \right) = 1 + \sum_{i=1}^N \ S_i \ 2^{i-1} \hskip1.0cm 1
\leq s \left( {\bf S} \right) \leq 2^N .
$$
So, the set of all possible $N$-Boolean function on $ N $ Boolean arguments \break $ {\bf F}: \mathbb{Z}_2^N \to \mathbb{Z}_2^N $ is isomorphic to the set of functional graphs on $ 2^N $ points $ {\cal G}_{2^N} $~${}^{4,5}$. See Fig.~1.

\begin{center}
\begin{picture}(150,134)
\thicklines
\put(-1,10){
\put(0,50){\circle*{5}}\put(0,80){\circle*{5}}\put(30,80){\circle*{5}}
\put(60,40){\circle*{5}}
\put(70,120){\circle*{5}}\put(80,60){\circle*{5}}\put(80,90){\circle*{5}}
\put(100,0){\circle*{5}}\put(100,30){\circle*{5}}\put(142,90){\circle*{5}}
\put(142,60){\circle*{5}}\put(91,120){\circle*{5}}

\thinlines \put(0,50){\vector(1,1){27}}\put(0,80){\vector(1,0){26}}

\put(60,40){\vector(1,1){17}}

\put(-80,40){\vector(1,1){27}} \put(-50,69){\circle*{5}} \put(-80,40){\circle*{5}}

\put(-50,70){\vector(1,0){27}} \put(-20,69){\circle*{5}}

\put(-20,70){\vector(0,-1){27}} \put(-20,40){\circle*{5}}

\put(-20,40){\vector(-1,1){27}}

\put(70,120){\vector(1,-3){8.8}}\put(80,90){\vector(0,-1){27}}
\put(80,60){\vector(2,-3){17.5}}\put(100,0){\vector(0,1){27}}
\put(100,30){\vector(-4,1){36.5}}\put(154,78){\vector(0,1){0}}\put(130,72){\vector(0,-1){0}}
\put(30,87){\oval(10,12)}\put(27.5,93){\vector(-1,0){0}}
\put(142,75){\oval(24,30)}\put(91,120){\vector(-1,-3){8.8}} \small
{\sf \put(1,40){\makebox(0,0){7}}\put(0,90){\makebox(0,0){5}}

\put(-80,50) {\makebox(0,0){1}}

\put(-50,80) {\makebox(0,0){2}}

\put(-20,80) {\makebox(0,0){3}}

\put(-20,30) {\makebox(0,0){4}}

\put(36,72){\makebox(0,0){6}}
\put(51,40){\makebox(0,0){12}}\put(61,120){\makebox(0,0){8}}
\put(90,63){\makebox(0,0){11}}\put(71,89){\makebox(0,0){10}}
\put(91,1){\makebox(0,0){14}}\put(110,33){\makebox(0,0){13}}
\put(143,100){\makebox(0,0){15}}\put(103,119){\makebox(0,0){9}}
\put(142,50){\makebox(0,0){16}} } }
\end{picture}

{\footnotesize{\bf Figure~1:} A functional graph {\it g } with $ n = 2^3 = 16 $. It is an element of $ {\cal G}_{2^3} $, and has four disconnected components whose cycles' lengths are (from left to right) {\bf 3}, {\bf 1}, {\bf 3}, and {\bf 2}.}

\end{center}

Note, however, that (1) defines a special type of Boolean endomorphism $  \mathbb{Z}_2^N \to  \mathbb{Z}_2^N $ due to the restrictions in the number of arguments and connections; as explained above. So, in general the \textit{NK}-Kauffman networks are not isomorphic to $ {\cal G}_{2^N} $.

Let $ \Psi $ be the function that associates to each Kauffman network $ \mathfrak{B} \in {\cal L}^N_K $, its corresponding functional graph $ g \in {\cal G}_{2^N} $, that is
$$
\Psi : {\cal L}^N_K \longrightarrow {\cal G}_{2^N} . \eqno(3)
$$
\textsf{Only} for the case $ K = N $, $ \Psi $ is a bijection: so $ {\cal L}^N_N \cong {\cal G}_{2^N} $~${}^{6,7}$.

Among the problems that scientists try to understand since 1969 is the asymptotic behaviour for $ N \gg 1 $, of the \textsf{average length} and the \textsf{average number of attractors} that Kauffman networks exhibit in terms of the parameters $ N $, $ K $, and $ p $. While there exists a general knowledge based on simulations, only two extreme cases have been addressed exactly, and also asymptotically: The case $ K = N $ with $ p = 1/2 $, also called {\it the random map model}~${}^{4,5,8}$. And the case $ K = 1 $ for $ p = 1/2 $~${}^{9}$. For general values of the parameters the best approach, to our knowledge, is a mean field approximation treatment done by Derrida {\it et al.} in 1986, showing a phase transition which divides the phase space in two zones where the behaviour of the networks is completely different~${}^{10}$. The phase transition curve is given by
$$
K_c \ 2 \, p_c \left( 1 - p_c \right) = 1 ,\eqno(4)
$$
see Fig.~2. The main result is that for values of $ K $ and $ p $ such that $ K < K_c $; two different states, which differ in $ d \ll N $ Boolean variables $ S_i $ will converge to the same state exponentially. Instead, for $ K > K_c $, they will diverge exponentially.

\begin{figure}

\centering

\includegraphics{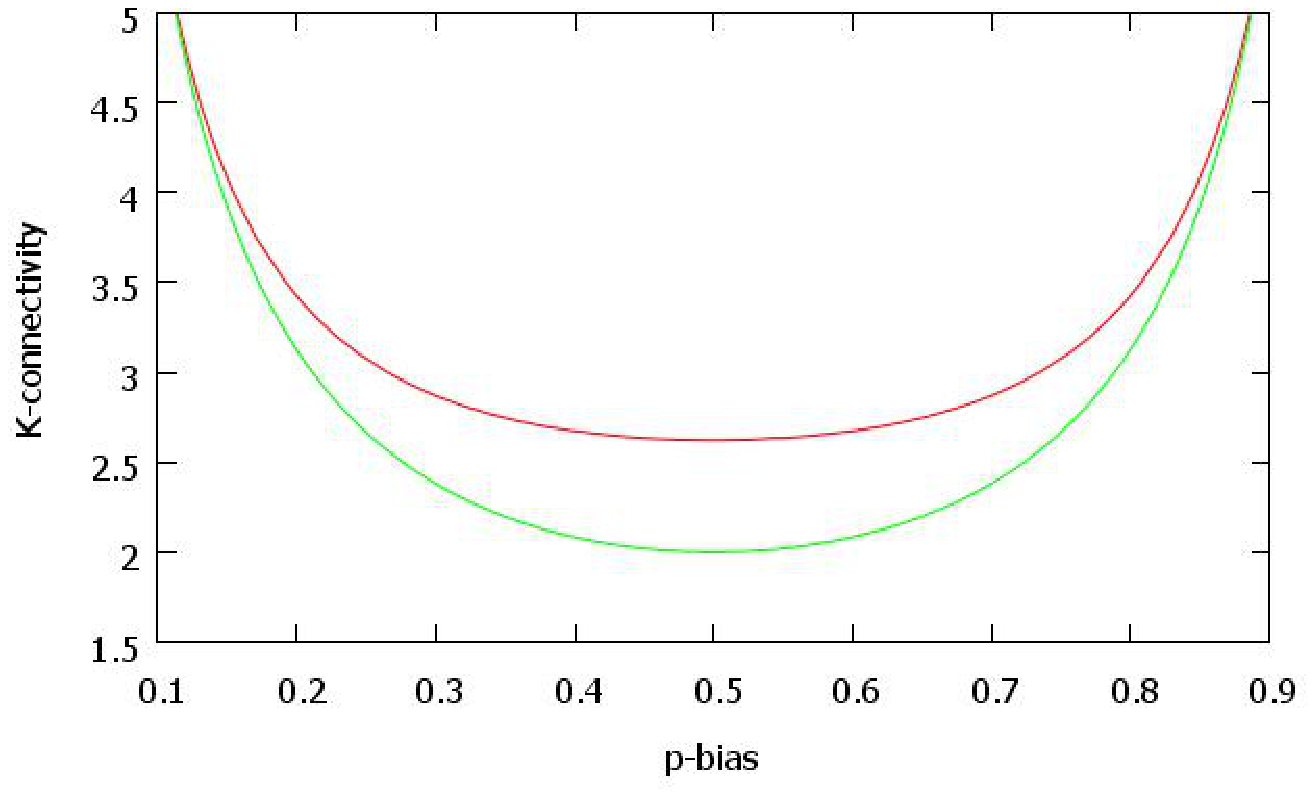}

\caption{{Figure~2.} \emph{The mean field approximation for the phase transition.}The phase transition of the green (the lower) curve corresponds to the Derrida {\it et al.} curve Eq.~(4), and the red one (the upper) to the curve (9) obtained considering the degree of irreducibility of Boolean functions.}

\label{locoton}

\end{figure}

The mechanism that regulates the phase transition at the level of the behaviour the Boolean functions $ b_K^{(i)} $ of (1) have long been discussed and mainly attributed to a type of Boolean function called \textsf{canalizing, or forcing}~${}^{2,3,11}$. In the following we are going to show that this is not the case. Instead, we are going to introduce a classification of Boolean functions called \textsf{degree of irreducibility}~${}^{7}$. We will show that this classification explains the phase transition, and also displaces the position of the Derrida {\it et al.} curve (4) upwards~${}^{12}$. See Fig.~2. Even more, the irreducible classification is responsible of the injective properties of (3), as well as the stability of Kauffman networks, against random changes of Boolean functions and connections~${}^{6,7}$.

This work is organized as follows: In Sec.~2 we establish the mathematical concepts necessarily to follow it. Sec.~3 introduces mathematically the concept of {\it irreducible Boolean classification}. Sec.~4 explains the random way by which Boolean functions are extracted, and de concept of {\it canalizing Boolean functions} is formally introduced. Sec.~5 contains a review of the mean field treatment, corrected for {\it irreducibility of Boolean functions}, to study the phase transition curve of \textit{NK}-Kauffman networks. This curve separates phase space into a ordered phase where small changes of the state do not matter; to a disordered phase where they give rise to big changes in the dynamics. Also {\it irreducible} vs. {\it canalizing} Boolean functions are counted. It is concluded that canalization cannot explain the phase transition curve due to its overwhelming minority in relation to {\it totally irreducible} Boolean functions for big values of $ K $. In Sec.~6 we see that boolean irreducibility determines de injectivity of $ \Psi $ through a critical connectivity $ \hat{K} $. In Sec.~7 it is seen that boolean irreducibility is also responsible of the stability of Kauffman networks against random changes of their connections. In Sec.~8 we establish our conclusions.


\section{2. Mathematical framework.}

Let's go into details:

Equation (1) is a set of $ N $ functions of the form
$$
\mathbb{Z}_2^N \ \stackrel{\ C_K^{*(i)}}{\longrightarrow} \ \mathbb{Z}_2^K \ \stackrel{\ b_K^{(i)}}{\longrightarrow} \ \mathbb{Z}_2 \hskip1.0cm i = 1, \dots, N; \eqno(5)
$$
where each function $ C_K^{*(i)} \left( S_1, \ldots S_N \right) = \left( S_{i_1}, \ldots S_{i_K} \right) $ is randomly constructed with equiprobability. It selects $K$ of the $ N $ variables, while each function $ b_K^{(i)} $ is termed a $K$-Boolean function.

Any $K$-Boolean function $ b_K^{(i)} $ is completely determined by its truth table, where for each of the $ 2^K $ possible inputs $ {\bf S} \in \mathbb{Z}_2^K $ a corresponding output $ \sigma_s \in \mathbb{Z}_2 $ ($ 1 \leq s \leq 2^K $) is obtained; so, there are a total of $ 2^{2^K} $ $K$-Boolean functions whose set we denote by $ \Xi_K $. See \textsf{Table~1} for an example of a truth table of a $ b_K^{(i)} $.

\footnotesize
\begin{center}
\begin{tabular}{|c|c|c|c|c|c|c|c|c|c|}
\hline $ S_1  $ & $ S_2 $  & $ S_3 $ & $ S_{..} $ & $ S_{..} $ & $ S_{..} $  & $ S_{K - 2} $ & $ S_{K - 1} $ & $ S_K $ & $ b_K^{(i)} $  \\
\hline $ 0 $ & 0 & 0 & 0 & 0 & 0 & 0 & 0 & 0 & $ \sigma_1 $ \\
\hline $ 1 $ & 0 & 0 & 0 & 0 & 0 & 0 & 0 & 0 & $ \sigma_2 $ \\
\hline $ 0 $ & 1 & 0 & 0 & 0 & 0 & 0 & 0 & 0 & $ \sigma_3 $ \\
\hline $ \cdots $ & $ \cdots $ & $ \cdots $ & $ \cdots $ & $ \cdots $ & $ \cdots $ & $ \cdots $ & $ \cdots $ & $ \cdots $ & $ \sigma_{..} $ \\
\hline $ \cdots $ & $ \cdots $ & $ \cdots $ & $ \cdots $ & $ \cdots $ & $ \cdots $ & $ \cdots $ & $ \cdots $ & $ \cdots $ & $ \sigma_{..} $ \\
\hline $ \cdots $ & $ \cdots $ & $ \cdots $ & $ \cdots $ & $ \cdots $ & $ \cdots $ & $ \cdots $ & $ \cdots $ & $ \cdots $ & $ \sigma_{..} $ \\
\hline $ 1 $ & 1 & 1 & $ \cdots  $ & $ \cdots $ & $ \cdots $ & 1 & 1 & 0 & $ \sigma_{2^K-2} $ \\
\hline $ 1 $ & 1 & 1 & $ \cdots $ & $ \cdots $ & $ \cdots $ & 1 & 0 & 1 & $ \sigma_{2^K-1} $ \\
\hline $ 1 $ & 1 & 1 & $ \cdots $ & $ \cdots $ & $ \cdots $ & 1 & 1 & 1 & $ \sigma_{2^K} $ \\
\hline

\end{tabular}
\end{center}
\normalsize \baselineskip = 12.4pt

\noindent
\textbf{Table~{\it 1}.} {\bf The general form of a truth table of a $K$-Boolean function.}

\noindent
\textsf{The rows correspond to the $ 2^K $ possible inputs of $ b_K^{(i)}$ and their outputs $ \sigma_s $, ($ 1 \leq s \leq 2^K $). Note that one can form a total of $ 2^{2^K} $ outputs which corresponds to the total possible number of $K$-Boolean functions, {\it i.e.} $ \# \Xi_K = 2^{2^K} $.}

\

\footnotesize
\begin{center}
\begin{tabular}{|c|c|c|c|c|c|c|c|c|c|c|c|c|c|c|c|c|}
\hline $ S_1 \, S_2 $ & $ \neg \tau $  & $\neg \vee$ & $ \nRightarrow $  & $ \neg S_2 $  & $ \nLeftarrow $ & $ \neg S_1 $ & $ \nLeftrightarrow $ & $ \neg \wedge $ & $ \wedge $  & $ \Leftrightarrow $ & $ S_1 $ & $ \Leftarrow $ & $ S_2 $ & $ \Rightarrow $ & $ \vee $ & $ \tau $  \\
\hline $ 0 \ 0  $ & 0 & 1 & 0 & 1 & 0 & 1 & 0 & 1 & 0 & 1 & 0 & 1 & 0 & 1 & 0 & 1 \\
\hline $ 1 \ 0  $ & 0 & 0 & 1 & 1 & 0 & 0 & 1 & 1 & 0 & 0 & 1 & 1 & 0 & 0 & 1 & 1 \\
\hline $ 0 \ 1  $ & 0 & 0 & 0 & 0 & 1 & 1 & 1 & 1 & 0 & 0 & 0 & 0 & 1 & 1 & 1 & 1 \\
\hline $ 1 \ 1  $ & 0 & 0 & 0 & 0 & 0 & 0 & 0 & 0 & 1 & 1 & 1 & 1 & 1 & 1 & 1 & 1 \\
\hline $ \lambda $ & 0 & 2 & 2 & 1 & 2 & 1 & 2 & 2 & 2 & 2 & 1 & 2 & 1 & 2 & 2 & 0 \\ \hline $ \omega $ & 0 & 1 & 1 & 2 & 1 & 2 & 2 & 3 & 1 & 2 & 2 & 3 & 2 & 3 & 3 & 4 \\
\hline

\end{tabular}
\end{center}
\normalsize \baselineskip = 12.4pt

\textbf{Table~{\it 2}.} {\bf The $ {\bf 16} $ truth tables of the sixteen $2$-Boolean functions.}

\noindent
\textsf{The first row corresponds to the four possible inputs of the functions, their {\it degree of irreducibility} $ \lambda $, and their {\it weight} $ \omega $. The symbols $ \neg $ and $ / $ stand for the negation of Boolean functions and variables. They change `zeros' to `ones' and viceversa.}

\

Before we move on into the mathematics; let us consider the truth table for the case $ K = 2 $ depicted in \textsf{Table~2}, here there are $ 2^{2^2} = 16 $ functions. At the first row the inputs $ S_1, S_2 $, and the logical meaning of each function is depicted. Their outputs are depicted along the next four rows. The last two rows correspond to their degree of irreducibility $ \lambda $, and their weight $ \omega $; to be explained below. We denote by $ \neg $ or $ / $ the negation of each Boolean variable or $K$-Boolean function; whose meaning is just, change a zero for a one and viceversa; {\it i.e.} $ \neg 1 = 0 $ and $ \neg 0 = 1 $. Note that the first eight functions of \textsf{Table~2} are the negations of the last eight. From the figure one can see that: {\it i)} The two functions {\it tautology} $ = \tau $, and {\it contradiction} $ = \neg \tau $ do not depend on neither of the arguments $ S_1 $ and $ S_2 $; since they have constant values. {\it ii)} The four functions $ S_i $, $ \neg \, S_i $ ($ i = 1, 2 $) only depend on one argument. {\it iii)} The remaining ten functions depend on both arguments. Let us generalise these concepts for any value of $ K $.

\section{3. The irreducible Boolean classification.}

\begin{itemize}

\item[{\it i})] A $K$-Boolean function $ b_K^{(i)} $ is {\it irreducible} on its $m$-th argument $ S_m $ ($ m = 1, \dots, K $), iff there exists an $ {\bf S} \in \mathbb{Z}_2^K $ for which
$$
b_K^{(i)} \left( S_1, \dots, \neg \, S_m, \dots, S_K \right) = \neg \ b_K^{(i)} \left( S_1, \dots, S_m, \dots, S_K \right),
$$
while, if this does not happen, the $K$-Boolean function $ b_K^{(i)} $ is {\it reducible} on the $m$-th argument $ S_m $.

\item[{\it ii})] A $K$-Boolean function $ b_K^{(i)} $ is said to have a {\it degree of irreducibility} $ \lambda $, with $ \lambda = 0, 1, \dots, K $; if it is irreducible on $ \lambda $ of their arguments and reducible on the remaining $ K - \lambda $.

\item[{\it iii})] If $ \lambda = K $, the $K$-Boolean function is called to be {\it totally irreducible}.

\end{itemize}

We denote by $ \lambda \left( b_K^{(i)} \right) $, the function that gives the {\it degree of irreducibility} of $ b_K^{(i)} $, and by $ \lambda $ ($ 0 \leq \lambda \leq K $) their possible values. Then, the sets
$$
{\mathfrak T}_K \left( \lambda \right) = \left\{ b_K^{(i)} \in \Xi_K | \lambda\left( b_K^{(i)} \right) = \lambda \right\}
$$
are disjoint, and cover $ \Xi_K $
$$
\Xi_K = \bigcup_{\lambda = 0}^K \ {\mathfrak T}_K \left( \lambda
\right), \eqno(6)
$$
The cardinalities $ \beta_K \left( \lambda \right) \equiv \# {\mathfrak T}_K \left( \lambda \right) $ have been calculated, with the result~${}^{7,12}$
$$
\beta_K \left( \lambda \right) = {K \choose \lambda} \ \sum_{m = 0}^\lambda \left( -1 \right)^{\lambda - m} \, {\lambda \choose m} \ 2^{2^m} . \eqno(7)
$$

\section{4. The canalization of Boolean functions.}

For many years the stabilisation of a Boolean function has been attributed to the concept of {\it canalization}${}^{11}$. This is, when the function is canalizing it gives rise to very short cycle dynamics~${}^{2,3,13}$. This has been shown to be truth when the network consists of one, and the same $ b_K^{(i)} $ and the inputs are $K$ nearest neighbour  on a linear array of $N$ inputs~${}^{2,3}$. So, no random selection of the inputs an the functions have been done. This gives, however, no guaranty that canalization stabilises the dynamics when random construction of the automata is done. We are going to show that this is untenable for the random construction done for Kauffman networks. First of all, from the probability (2), that the value each the $ 2^K $ outputs of $ b_K^{(i)} $, $ \left\{ \sigma_1, \ldots, \sigma_{2^K} \right\} $, be  $ \sigma_s = 1 $; follows that the probability of extraction of each $ b_K^{(i)} $ is~${}^{7,12}$
$$
\Pi_p \left( b_K^{(i)} \right) = \Pi_p \circ \omega \left( b_K^{(i)} \right) = p^\omega \left( 1 - p \right)^{2^K - \,
\omega} \ , \eqno(8)
$$
where $ \omega = 0, 1, \dots, 2^K $ is the value of the {\it weight function} $ \omega \left( b_K^{(i)} \right) $, defined by,
$$
\omega \left( b_K^{(i)} \right) = \sum_{s=1}^{2^K} \sigma_s .
$$

Let us now go to canalization: a $ b_K^{(i)} $ function is canalizing iff there exists at least an input $ S_\alpha $ ($1\leq \alpha \leq K$), and $ \xi, \tau \in \mathbb{Z}_2 $ such that
$$
S_\alpha = \xi \ \ \Rightarrow \ \ b_K^{(i)} \left( S_1, \dots, S_{\alpha-1}, \xi, S_{\alpha+1}, \dots, S_K \right) = \tau ,
$$
for all $ S_\beta $ with $ \beta \neq \alpha $~${}^{13,14}$. Note that the canalization (or forcing) condition sets drastically the value of $ b_K^{(i)} $ to $ \tau $ once the canalizing variable attains its canalizing value $ \xi $. When the network is composed of the very same canalizing functions with their inputs on a lattice, the canalizing condition tends to prevail and the network settles on a short cycle composed mainly of $ \tau $'s~${}^{2,13}$.

From \textsf{Table~2} one can see that for $ K = 2 $ all but the functions $ \Leftrightarrow $ and $ \nLeftrightarrow $ are canalizing; this has been taken as an explanation of the transition curve (4) in the region  $ p \sim 1/2 $~${}^{13}$. However in the regions $ p \sim 0 $ and $ p \sim 1 $, $ K_c \gg 1 $ the ratio of canalizing functions in relation to $ \# \, \Xi_K = 2^{2^K} $, decreases like $ 4 K / 2^{2^{K-1}} $~${}^{13}$. That is, canalizing functions constitute a minority with respect the total $K$-Boolean functions. In contrast, for $ K_c \gg 1 $ the ratio of the totally irreducible functions with respect to $ \# \, \Xi_K $ goes like $ 1 - {\cal O} \left( {K \over 2^{2^{K-1}}}\right) $; so, almost any $K$-Boolean function is totally irreducible with respect to the counting measure~${}^{12}$. The only way that canalization could play a role would be if canalizing functions have better chances to be extracted in the regions $ p \sim 0 $ and $ p \sim 1 $. This could happen due that it is known that canalizing functions use to have many `0' or `1' in their outputs and so, from (8), better chances to be extracted~${}^{13}$. We are going to show below that this is not the case. We want to stress, that {\it canalization} and {\it irreducibility} are no related concepts~${}^{14}$. The reader may found two examples in \textsf{Table~3}.

\footnotesize
\begin{center}
\begin{tabular}{|c|c|c|c|c|}
\hline $ S_1  $ & $ S_2 $  & $ S_3 $ & $ b_3 (\alpha) $ & $ b_3 (\beta) $ \\
\hline  0 & 0 & 0         & 0 & 0 \\
\hline  1 & 0 & 0         & 0 & 1 \\
\hline  0 & 1 & 0         & 0 & 1 \\
\hline  1 & 1 & 0         & 0 & 0 \\
\hline  0 & 0 & 1         & 0 & 1 \\
\hline  1 & 0 & 1         & 1 & 0 \\
\hline  0 & 1 & 1         & 1 & 0 \\
\hline  1 & 1 & 1         & 0 & 1 \\
\hline

\end{tabular}
\end{center}
\normalsize \baselineskip = 12.4pt

\noindent
\textbf{Table~{\it 3}.} {\bf Canalization and Irreducibility for two $ K = 3 $ functions.}

\noindent
\textsf{The two functions are totally irreducible ($ \lambda = 3 $), however $ b_3 (\alpha) $ is canalizing since $ S_3 = 0 $ implies $ b_3 (\alpha) = 0 $ whatsoever the values of $ S_1 $, and $ S_2 $. On the contrary $ b_3 (\beta) $ is not canalizing. The operational representation of $ b_3 (\alpha) $, and $ b_3 (\beta) $ helps to clarify this situation: $ b_3 (\alpha) = \left( S_1 \oplus S_2 \right) \cdot S_3 $, while \break $ b_3 (\beta) = S_1 \oplus S_2 \oplus S_3 $; where $ \oplus $ means addition {\it modulo 2}.}

\section{5. Irreducibility vs. Canalization in the phase transition curve.}

Let us see now in which way irreducible classification (6) affect the Derrida {\it et al.} medium field approximation~${}^{10}$. As shown in Ref.~[12], their approximation is equivalent to the equation
$$
\left< Con \left( b_K^{(i)} \right) \ P_c \left( b_K^{(i)} \right) \right> \ = \ 1 .
$$
Where $ Con \left( b_K^{(i)} \right) $ is the connectivity of each Boolean function, {\it i.e.} the number of arguments that affect each $ b_K^{(i)} $, and $ P_c \left( b_K^{(i)} \right) $ the probability that the output of $ b_K^{(i)} $ changes, due that one of the arguments changes. The average is taken over the probability (8). In their calculations Derrida {\it et al.} took $ Con \left( b_K^{(i)} \right) = K $, {\it i.e.} just the number of arguments that each $ b_K^{(i)} $ has. Then since $ \left< P_c \right> = 2 \ p \, \left( 1 - p \right) $; equation (4) is obtained~${}^{10}$. Instead taking account of decomposition (6) the \textsf{real connectivity} for each $ b_K^{(i)} $ should be $ \lambda \left( b_K^{(i)} \right) $. So, taking $ Con \left( b_K^{(i)} \right) = \lambda \left( b_K^{(i)} \right) $, we obtain~${}^{12}$
$$
2 \, K_c \, p_c \left( 1 - p_c \right) \left\{ 1 - 2 \, p_c \, \left( 1 - p_c \right) \left[ 1 - 2 \, p_c \, \left( 1 - p_c \right) \right]^{2^{K_c - 1}-2} \right\} = 1 . \eqno(9)
$$
Note that (4) is corrected by a new factor $ \left\{ \cdots \right\} \leq 1 $, and that $ \left\{ \cdots \right\} \sim 1 $ in the regions $ p \sim 0 $ and $ p \sim 1 $. So, the new curve approaches asymptotically the Derrida {\it et al.} curve in the extreme values of $ p $. See Fig.~2 for a comparison of the two curves.

We then see, that the average $ \left< \lambda \left( b_K^{(i)} \right) \ P_c \left( b_K^{(i)} \right) \right> $ regulates the phase transition; and no reference has been done to canalization. Nevertheless, let us see the proportion of canalizing and totally irreducible functions along the curve (9). Just, Shmulevich \& Konvalina in 2004 calculated the probability $ \Pi_p \left( \mathfrak{C}_K \right) $ that under extraction (8) a function $ b_K^{(i)} $ belongs to the set of canalizing functions $ \mathfrak{C}_K $ (equation (19) of Ref.~[13]). Now, we are going to calculate the probability $ \Pi_p \left( \mathfrak{T}_K \left( \lambda \right) \right) $, that under extraction $ b_K^{(i)} $ belongs to the set of $K$-Boolean functions with a degree of irreducibility $ \lambda $, {\it i.e.} $ \mathfrak{T}_K \left( \lambda \right) $. For that scope, we use the calculation that we did in Ref.~[14] for the number of $ b_K^{(i)} \in \Xi_K $ which have irreducible degree $ \lambda $ and weight $ \omega $; denoted by $ \varrho_K \left( \lambda, \omega \right) $, there we obtained:
\begin{eqnarray*}
\varrho_K \left( \lambda, \omega \right) &=& {K \choose \lambda} \ \sum_{m=0}^\lambda \left( -1 \right)^{\lambda - m} {\lambda \choose m} \\ \, & \times & \delta \left( \left\lfloor \omega \, 2^{m-K} \right\rfloor - \omega \, 2^{m-K} \right) \ \ {2^m \choose \left\lfloor \omega \, 2^{m-K} \right\rfloor} ,
\end{eqnarray*}
where for all $ a \in \mathbb{R} $,

$$
\delta \left( a \right) = \left\{ \begin{array}{ll}
1 & \mbox{if $ a = 0 $}
\\ {} & {} \\ 0 & \mbox{if $ a \neq 0 $} \end{array} \right.
$$
is Kronecker's delta, $ \lfloor a \rfloor \in \mathbb{Z} $ the {\it floor function}, which denotes the greatest integer $ \lfloor a \rfloor $ such that $ \lfloor a \rfloor \leq a $, $ {\lambda \choose m} = 0 $, for $ m > \lambda $, and $ 0^0 \equiv 1 $.

Then from (8), $ \Pi_p \left( \mathfrak{T}_K \left( \lambda \right) \right) $ is given by
\begin{eqnarray*}
\Pi_p \left( \mathfrak{T}_K \left( \lambda \right) \right) &=& \sum_{\omega=0}^{2^K}  \varrho_K \left( \lambda, \omega \right) \ \Pi_p \left( \omega \right) \\ &=& {K \choose \lambda} \sum_{m=0}^\lambda \left( -1 \right)^{m-\lambda} {\lambda \choose m} \left[ p^{2^{K-m}} + \left( 1 - p \right)^{2^{K-m}} \right]^{2^m} . \hskip0.75cm (10)
\end{eqnarray*}
Now we may calculate along the phase transition curve (9) the average numbers of totally irreducible functions, and canalizing functions. The results are shown in \textsf{Table~4} and imply  that canalization plays no role at all. Instead there is a predominance of the functions in $ \mathfrak{T}_K \left( K \right) $ (the totally irreducible) as the values of $ p $ go to the extremes $ p \sim 0 $ and $ p \sim 1 $. We may conclude that for random constructed automata, as Kauffman's networks, is not canalization which determines their dynamics.

\footnotesize
\begin{center}
\begin{tabular}{|c|c|c|c|}
\hline $ p  $ & $ K_c $  & $ \Pi_p \left( \mathfrak{C}_K \right) $ & $ \Pi_p \left( \mathfrak{T}_K \left( K \right) \right) $ \\

\hline  0.73460 & 3 & $ 0.71784  $ & 0.76274 \\
\hline  0.84562 & 4 & $ 0.71715  $ & 0.95574 \\
\hline  0.88618 & 5 & $ 0.56228  $ & 0.99324 \\
\hline  0.90818 & 6 & $ 0.28972   $ & 0.99309 \\
\hline  0.92258 & 7 & $ 6.0099 \times 10^{-2} $ & 0.99984 \\
\hline  0.93301 & 8 & $ 2.0986 \times 10^{-3} $ & 0.99999 \\
\hline  0.94096 & 9 & $ 3.0759 \times 10^{-6} $ & 0.99999 \\
\hline

\end{tabular}
\end{center}
\normalsize \baselineskip = 12.4pt

\noindent
\textbf{Table~{\it 4}.} {\bf Canalization {\it vs.} Irreducibility along the phase transition.}

\noindent
\textsf{The probability of extraction of canalizing functions $ \mathfrak{C}_K $ and totally irreducible functions $ \mathfrak{T}_K \left( K \right) $ along the phase transition curve $ \left( p, K_c \right) $ given by (9). Only values of $ p $ with $ p > 1/2 $ are considered since the curve is symmetrical at $ p = 1 / 2 $.}

\

\section{6. The effect of irreducibility in the injectivity of $ \Psi $.}

Decomposition (6), determines, also, the injective properties of function $ \Psi $ defined by (3), that may have consequences to the average dynamics of Kauffman networks in a way that we still do not know. Let $ \vartheta \left( N, K \right) $ be the average number of Kauffman networks that $ \Psi $ maps onto the same functional graph $ g $ for the \textsf{equiprobability} case $ p = 1 / 2 $. Then, we have shown that in the asymptotic regime $ N \gg 1 $ exists a critical connectivity $ \hat{K} $ given by
$$
\hat{K} \approx \log_2 \log_2 \left( {2 N \over \ln 2} \right) + {\cal
O} \left( {1 \over N \ln N } \right); \eqno(11)
$$
such that $ \vartheta \left( N, K \right) \approx e^{\varphi \, N} \gg 1 $ ($ \varphi > 0 $) or $ \vartheta \left( N,K \right) \approx 1 $, depending on whether $ K < \hat{K} $ or $ K > \hat{K} $, respectively~${}^{6,7}$. Note that this result comes as a surprise if we compare it, with the phase transition curve (9) where no trace of $ N $ is left. We hypothesize  that this has to be with the nature of each calculation: The phase transition curve is obtained through a mean field approximation where $ N \to \infty $ at the end of it. On the contrary, the calculation of the injective transition regime (11) is {\bf exact}. The asymptotic approximation involved, is only used because the relevant term becomes (11) very soon as $ N $ grows while the other terms are of no profit. Even for $ N = 10 $ the relative error is inferior to $ 5 \% $ as one may see in Fig.~4 of Ref.~[6]. so the other terms almost do not change the result. The exact expressions can be consulted in the Appendix of Ref.~[7].

\section{7. Irreducibility and Stability.}

Decomposition (6) also accounts for the stability of Kauffman networks against random changes of their connections $ C_K^{*(i)} $. The probability $ {\cal P} \left( {\cal A} \right) $ that (5) remains invariant under a random change $ C_K^{*(i)} \to \bar{C}_K^{*(i)} $ were calculated in equations (23) of Ref.~[7] with the result (adapted to the present notation):
$$
{\cal P} \left( {\cal A} \right) = \sum_{\lambda=0}^K \ { K! \, \left( N - \lambda \right)! \over N! \, \left( K - \lambda \right)! } \ \Pi_p \left( \mathfrak{T}_K \left( \lambda \right) \right) \eqno(12)
$$
where $ \Pi_p \left( \mathfrak{T}_K \left( \lambda \right) \right) $ is given by (10). Once again since the calculation were done for $ N \gg 1 $, and the first term of the addition prevails, making calculation easier. However the point that we want to stress from (12) is the fact that, is the whole addition on $ \lambda $, and so decomposition (6) with their cardinalities (7) that determine the stability properties of Kauffman networks.

\section{8. Conclusion.}

Equations (9), (10), (11), and (12) are unavoidable traces that the irreducible decomposition (6) and the combinatorial expressions that are obtained from it; play an analytical role in the study of the average dynamics of the \textit{NK}-Kauffman networks when arbitrary values of $N$, $K$, and $p$ are considered. In particular, we have shown that canalization could not determine the dynamics of Kauffman's networks as long as canalizing functions are very few in relation to totally irreducible functions along the phase transition curve (9). See Table~4.

\section{Acknowledgments}

This work was supported by {\bf PAPIIT} project No.~{\bf IN102712-3}. The author thanks: \textsf{Alberto Verjovsky} for critical reading of the manuscript, \textsf{Pilar L\'opez Rico} for informatics services, \textsf{Thal\'\i a Figueras} for careful suggestions to improve the manuscript, \textsf{V\'\i ctor Dom\'\i ngez} and \textsf{Fernando Gonz\'alez} for computer's advise. Last but not least, the author has benefit from fruitful discussions with \textsf{Fabio Benatti} in his stay at {\it Universit\`a degli Studi di Trieste, Italy}, where part of this work was done.

\


{\bf References}

\begin{itemize}

\item[${1}$] Kauffman, S.A.; {\rm Metabolic stability and epigenesis in randomly connected Nets}. {\it J.~Theoret.~Biol.} {\bf 22} (1969) 437.

\item[${2}$] Kauffman, S.A.; {\rm The Origins of Order: Self-Organization and Selection in Evolution}. Oxford University Press (1993).

\item[${3}$] Aldana, M., Coppersmith, S. and Kadanoff, L.; {\rm Boolean dynamics with random couplings}. In: Perspectives and Problems in Nonlinear Science, 23--89. Springer Verlag, New York (2003).

\item[${4}$] Romero, D. and Zertuche, F.; {\rm The asymptotic number of attractors in the random map model}. {\it J.~Phys.~A:~Math.~Gen.} {\bf 36} (2003) 3691.

\item[${5}$] Romero, D. and Zertuche, F.; {\rm Grasping the connectivity of random functional graphs}. {\it Stud.~Sci.~Math.~Hung.} {\bf 42} (2005) 1.

\item[${6}$] Romero, D. and Zertuche, F.; {\rm Number of different binary functions generated by \textit{NK}-Kauffman networks and the emergence of genetic robustness}. {\it J.~Math.~Phys.} {\bf 48} (2007) 083506.

\item[${7}$] Zertuche, F.; {\rm On the robustness of NK-Kauffman networks against changes in their connections and Boolean functions}. {\it J.~Math.~Phys.} {\bf 50} (2009) 043513.

\item[${8}$] Derrida, B. and Flyvbjerg, H.; {\rm The random map model: a disordered model with deterministic dynamics}. {\it J.~Physique.} {\bf 48} (1987) 971.

\item[${9}$] Flyvbjerg, H. and Kjaer, N.J.; {\rm Exact solution of Kauffman's model with Connectivity One}. {\it J.~Phys.~A:~Math.~Gen.} {\bf 21} (1988) 1695.

\item[${10}$] Derrida, B. and Stauffer, D.; {\rm Phase transitions in two-dimensional cellular automata.} {\it Europhys.~Lett.} {\bf 2} (1986) 739;  Derrida, B. and   Pomeau, Y. {\rm Random Networks of Automata: A Simple Annealed Approximation}. Europhys.~Lett. {\bf 1} (1986) 45; Derrida, B. and  Weisbuch, G. {\rm Evolution of overlaps between configurations in random Boolean networks}. J.~Physique {\bf 47} (1986) 1297.

\item[${11}$] Kauffman, S.A.; {\rm The Large Scale Structure and Dynamics of Gene Control Circuits: An Ensemble Approach}. {\it J.~Theoret.~Biol.} {\bf 44} (1974) 167.

\item[${12}$] Zertuche, F.; {\rm Phase transition in NK-Kauffman networks and its correction for
Boolean irreducibility}. {\it Physica~D} {\bf 275} (2014) 35-42.

\item[${13}$] Just, W., Shmulevich, I. and Konvalina, J.; {\rm The number and probability of canalizing functions.} {\it Physica~D} {\bf 197} (2004) 211.

\item[${14}$] Takane,  M. and Zertuche, F.; {\rm $ \mathbb{Z}_2 $-Algebras in the Boolean Function Irreducible Decomposition}. {\it J.~Math.~Phys.} {\bf 53} (2012) 023516.

\end{itemize}

\end{document}